# Flexibility Analysis for Smart Grid Demand Response

Sarah O'Connell and Stefano Riverso
United Technologies Research Centre, Ireland Ltd.
Cork, Ireland
oconnesa@utrc.utc.com

*Abstract*—Flexibility is a key enabler for the smart grid, required to facilitate Demand Side Management (DSM) programs, managing electrical consumption to reduce peaks, balance renewable generation and provide ancillary services to the grid. Flexibility analysis is required to identify and quantify the available electrical load of a site or building which can be shed or increased in response to a DSM signal. A methodology for assessing flexibility is developed, based on flexibility formulations and optimization requirements. The methodology characterizes the loads, storage and on-site generation, incorporates site assessment using the ISO 50002:2014 energy audit standard and benchmarks performance against documented studies. An example application of the methodology is detailed using a pilot site demonstrator.

## I. INTRODUCTION

This paper defines flexibility, reviews optimization approaches, formulates a methodology for assessment and applies it to a pilot site.

Demand Side Management (DSM) incorporating Demand Response (DR) and ancillary services are increasingly being used as a means of balancing electricity grids globally. The European Commission's 2020 targets to generate 20% of Europe's energy from renewable energy, cut annual primary energy consumption by 20% and reduce GHG emissions by 20% [1] have already resulted in increased renewable generation, with installed capacities set to increase to meet the targets. Increasing penetration of renewables above 25% [1] requires increased flexibility to enable Transmission System Operators (TSOs) and Distribution System Operators (DSOs) balance non-dispatchable sources, avoid grid perturbation, manage the power locally to avoid transmission losses, reduce the installation of costly assets and overall ensure resilience and grid stability.

Markets for DSM which encourage the participation of aggregators have been present in the US for some time [2] [3], and are now expanding across Europe. For example, the UK's Demand Side Balancing Reserve (DSBR) has a minimum threshold of 1 MW but permits participation by a 3[rd] party intermediary (aggregator) offering DBSR from multiple sites [4]. France has gone a step further and is piloting a Block Exchange Notification of Demand Response (NEBEF) mechanism, with a minimum size of 0.1 MW, allowing greater direct participation for smaller buildings [5].

The ELSA (Energy Local Storage Advanced system) project [6], funded by the European Commission under the Horizon 2020 program, aims to implement and demonstrate the flexibility offered by the integration of distributed small to medium size storage systems, coupled with load management and local renewable generation to enable smart grid services. The storage systems consist of low cost second life electric vehicle Li-ion batteries, addressing the challenge of cost competitiveness, which has been identified as one of the four challenges for increased deployment of energy storage [7]. The renewable generation systems are Photovoltaic arrays (PV), installed on-site. Smart grid service use cases include demand response, peak shaving and ancillary services. A Building Energy Management System (BEMS) will manage the storage, loads and generation to provide these services locally and offer the flexibility to an aggregator or a DSO.

## II. FLEXIBILITY DEFINITION

The International Energy Agency's (IEA) Annex 67 project includes development of a standard definition for energy flexibility [8]. Annex 67 is due for completion in 2019, so in the interim the following definition of energy flexibility is proposed:

Modifying (decreasing or increasing) the electrical load profile through load shedding, ramping up, on site generation and storage, implemented using automatic control of systems, while minimizing the impact on occupants and operations.

## III. FORMULATION

At present, flexibility is assessed on a case-by- case basis. There is no standardized methodology and each site or aggregator relies on experts to determine what the flexibility for a specific site is. For single source applications, for example, running a backup generator, flexibility is easily assessed, but for sites where multiple sources of flexibility are proposed, determining what can be offered becomes more complex. Contracts with aggregators, DSOs and TSOs are based on committing to a defined range of flexibility. The site owner needs to know the flexibility range in order to select the most appropriate demand side management or flexibility program to participate in, or to decide if participation is a worthwhile option for them. Aggregators wish to minimize up front time and effort when assessing sites for aggregated portfolios. Having a fast, easily implementable and standardized assessment methodology is an enabler for both aggregators and site owners. This assessment methodology is done a priori, before the installation of an ICT platform hosting algorithms.

The formulation for assessing load flexibility [9], F, shown in (1)

$$F_{l,p}(t) = S_{l,p}(t) \cdot \min [C_{l,p}(t), A_{l,p}(t)] \quad (1)$$

has three components: sheddable, S, controllable, C, and acceptable, A. Subscripts l and p denote the type of load and type of product, respectively. Flexibility is expressed as a percentage of total load (%). The resource potential, R, may then be calculated from:

$$R_{l,p}(t) = F_{l,p}(t) \cdot L_l(t) \quad (2)$$

Under this definition, for a load to be flexible, it must be sheddable, controllable and acceptable. However, on-site generation using non-dispatchable renewables (e.g. wind or PV) has the capability to provide flexibility even though they are not controllable. On-site storage may not fit into this categorization also as it is not sheddable, but it has the capability to shift consumption to a different time period. Another consideration is that if the flexibility, F, of all loads are summed, this may not account for interactions between loads. Acceptability requires the end user in the loop, for example if the proposed load reduction impacts on comfort, it may be site specific.

Determining the shedability, level of control available and acceptability of shedding loads, or a proportion of that load, over the span of one 24 hour period is a non-trivial task. It requires access to data, development of a methodology or approach, energy audits and expert analysis.

The above formulation gives a high level view, however, for each load or source of flexibility, there are other factors which have to be taken into consideration. A further set of parameters are required to define the flexibility in more detail [10]. These include the amount of power to be increased or decreased, the duration of the action in hours, Time in Advance (TIA) notification in advance of the action, the extra power required before and after the flexibility action (also known as rebound or spring back). These parameters represent the characteristics or constraints of individual load flexibility in a more detailed way.

Another approach [11] incorporates flexibility activation constraints such as minimum and maximum number of activations and recovery between activations, but omits some of those included above [10].

A different way to categorize load flexibility is to identify the loads as classes [12]. Five classes of loads are identified: shiftable profile, shiftable volume, curtailable loads which consist of reducible and disconnectable loads, and finally, inflexible loads.

## IV. OPTIMISATION

Energy flexibility is essentially a scheduling problem, well suited to optimization. From the approaches outlined below, the optimization objective function requires a number of constraints, including but not limited to:
- Flexibility parameters
- Pricing & market constraints
- Resource Availability & Load Forecasts
- Grid and/or Aggregator signals e.g. OpenADR.

This paper addresses the flexibility parameter constraints.

Mixed Integer Linear Programming (MILP) was used [11] to formulate objective functions for three distinct flexibility scenarios: maximizing expected revenue, maximizing bid duration and maximizing peak power.

Maximizing expected revenue:

$$\text{Max } \{\sum_t \sum_r \text{Power}(t, r) \cdot \text{EnergyWeight}(t) \cdot u_t\} \quad (3)$$

where

$$\text{Power}(t, r) = \text{Res}(t, r) \cdot \text{EventImpact}(t, r); \quad (4)$$

Maximizing bid duration:

$$\text{Max } \{\sum_t \text{BidActive}(t)\} \quad (5)$$

Subject to the constraint:

$$\sum_r \text{Res}(t, r) \cdot \text{EventImpact}(t, r) \geq \text{BidActive}(t) \cdot \text{MinPower} \quad (6)$$

Maximizing peak power:

$$\text{Max } \{\text{Max}\sum_t \text{Res}(t, r) \cdot \text{EventImpact}(t, r), t \in T\}\} \quad (7)$$

Whereby, Power(t, r) is the power flexibility provided by the resource r for the timestep t and Res(t, r) is a binary variable indicating if resource r is active during timestep t; EventImpact(t, r) is the amount of power variation that can be provided by resource r at timestep t; EnergyWeight(t) is a unitless weighting factor related to electricity price; $u_t$ is the length of the optimization step; BidActive(t) is a binary variable indicating if the bid is active at time t; MinPower is the minimum power that must be provided; T is the set of time steps between the bid start and end times.

Flexibility in these objective functions is not explicitly included but is expressed as EventImpact. Applying a unitless weighting factor for price, while it keeps it independent of currency, adds a level of complexity that may not be useful. The output of the cost function would then have to be converted to a financial cost. Using Booleans to switch on or off flexibility sources is an approach which allows the overall flexibility to be adjusted according to changing source availability. It could be argued that maximizing bid duration and maximizing peak power would be the same as maximizing revenue, if the grid signals were price based. Maximizing peak power is interesting for a future scenario where renewable penetration has increased to the extent that grid operators require the load to increase in order to balance the grid.

A stochastic approach, also using MILP, is shown in (8) [12].

$$\min \sum_{s \in S} R_s \left[ \sum_{a \in A} \sum_{t \in T} P_{a,t,s}^{energy} x_{a,t,s}^{import} + \sum_{a \in A} P_a^{peak} x_{a,s}^{peak} + \right.$$

$$\sum_{o \in O} \sum_{y \in Y} \sum_{t \in T} G_{o,y}^{startup} \alpha_{o,y,t,s}^{start} + \sum_{d \in D^c} \sum_{y \in Y} \sum_{t \in T} X_{d,y} \varphi_{o,y,t,s}$$

$$\left. \sum_{a \in A} \sum_{t \in T} P_{a,t}^{sales} x_{a,t,s}^{export} \right]$$
(8)

The probability of scenarios is given by $R_s$. $P^{energy}$ denotes total load (kW) less flexibility (kW) while $x^{import}$ is the standard rate cost of energy (€). The next term includes for a peak premium, for example in a Critical Peak Pricing (CPP) use case. $G^{startup}$ relates to a converter start-up cost; the $X_d$ term incorporates losses due to disutility, such as loss of production or loss of worker productivity; while the final term includes for income from the sale of excess generation.

An advantage of the approach in (8) is that it is less tied to specific use cases than (3), (5) and (7). The pricing structure used in this cost function is well suited to Time of Use (TOU) or CPP fixed price structures, but may not be adaptable enough for Real Time Pricing (RTP) or intra-day demand response signals. Converter start up is not an issue with the battery, whereas it may be a consideration with on-site generators or fossil fuel power plants. Disutility is something one always seeks to avoid, but if there is a risk of it occurring, it is advisable to include it in the flexibility matrix. If a site has excess generation, there may be a feed in tariff in some sites, others may have spill to the grid with no payment while in some cases there may be a specific prohibition of net export.

## V. FLEXIBILITY CHARACTERISATION

The flexibility formulations outlined in Section III require complex assessment and expert analysis to produce the parameters required. This has to be done for every site considering participation in DSM. Deriving the parameters in a scalable and easily applicable way would overcome the need for building specific experts or continuous user on-line data gathering. To this end, a flexibility characterization process was developed.

The objective of the flexibility characterization process is the elimination of non-flexible loads and identification of flexibility matrix for storage, on-site generation and loads.

### A. Flexibility Characterisation

The flexibility characterization process is shown in Figure 1. An example of the application of the flexibility characterization process is as follows: An HVAC (Heating Ventilation & Air Conditioning) system is a load. It is possible to reduce the load over a short period, therefore it is sheddable, S. These are typically controlled by a Building Management System (BMS) so it is controllable, C. It must be determined if it is acceptable, A, to the building operator to reduce the HVAC power consumption such as by reducing or increase temperature set points or by displacing electrical heating by gas fired systems.

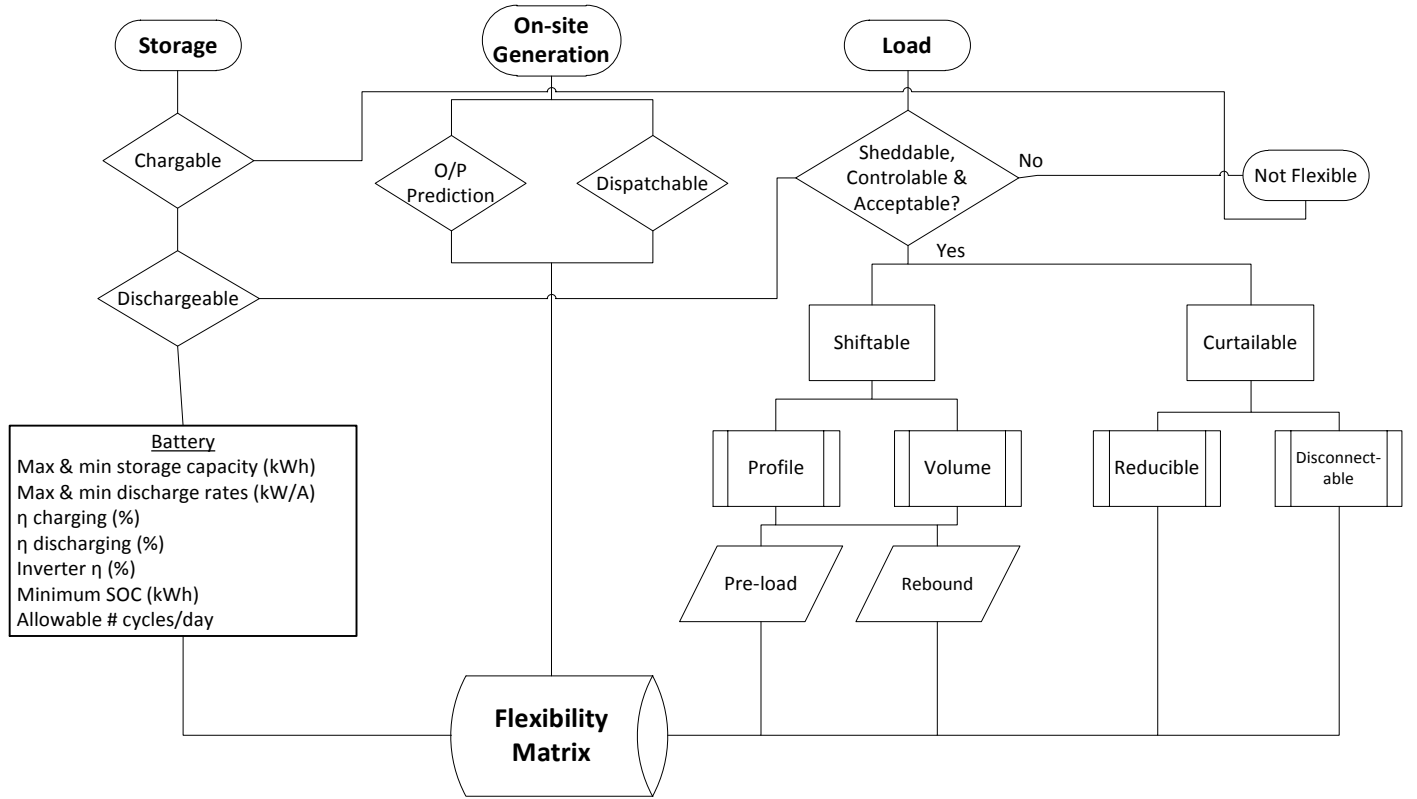

Fig. 1. Flexibility Characterization Process.

HVAC is generally a shiftable load, but it may be partly curtailable. E.g. if thermal energy is reduced during a flexibility event, there is often a rebound effect afterwards where more energy is needed to restore the building to the temperature set point. However, if the duration of the event is short, it may be possible to curtail the load with minimal impact on indoor air temperature and thereby avoid rebound.

The parameters gathered through this process are then input into the flexibility matrix, and the process is repeated for other loads, storage and on-site generation sources on site.

The flexibility matrix is a repository for constraints which are inputs to the building owner DSM program selection, aggregator or DSO site selection and algorithms for BEMS. The parameters in the flexibility matrix may include flexible power (kW), duration of event [h], TIA notification [h], pre-load (P, t) [kW, h], rebound (P, t) [kW, h], load availability (days, h), disutility cost, financial or other, shed time (s/min) and time frame when requests are permitted.

### B. Energy Audit

A detailed energy audit is required to evaluate the power or energy flexibility of a site [13]. In this paper, an energy audit is proposed as a systematic way of identifying building systems, quantifying loads, storage and sources and from that, assessing flexibility.

There are a number of standardized audit procedures such as IEA Annex 11 and ASHRAE [14]. Audits are categorized as Level 1, 2 or 3. Level 1 is a walkthrough, Level 2 is a standard audit and Level 3 is an investment grade audit. The international standard ISO 50002:2014 [15] contains a minimum set of specifications for energy audits, categorizing them as Type 1, 2 or 3, equivalent to Level 1, 2 and 3 above. The objective of the energy audit for this project is determining energy flexibility rather than identification of energy conservation measures, which energy audits are more typically used for. Some of the processes have been adapted to reflect this.

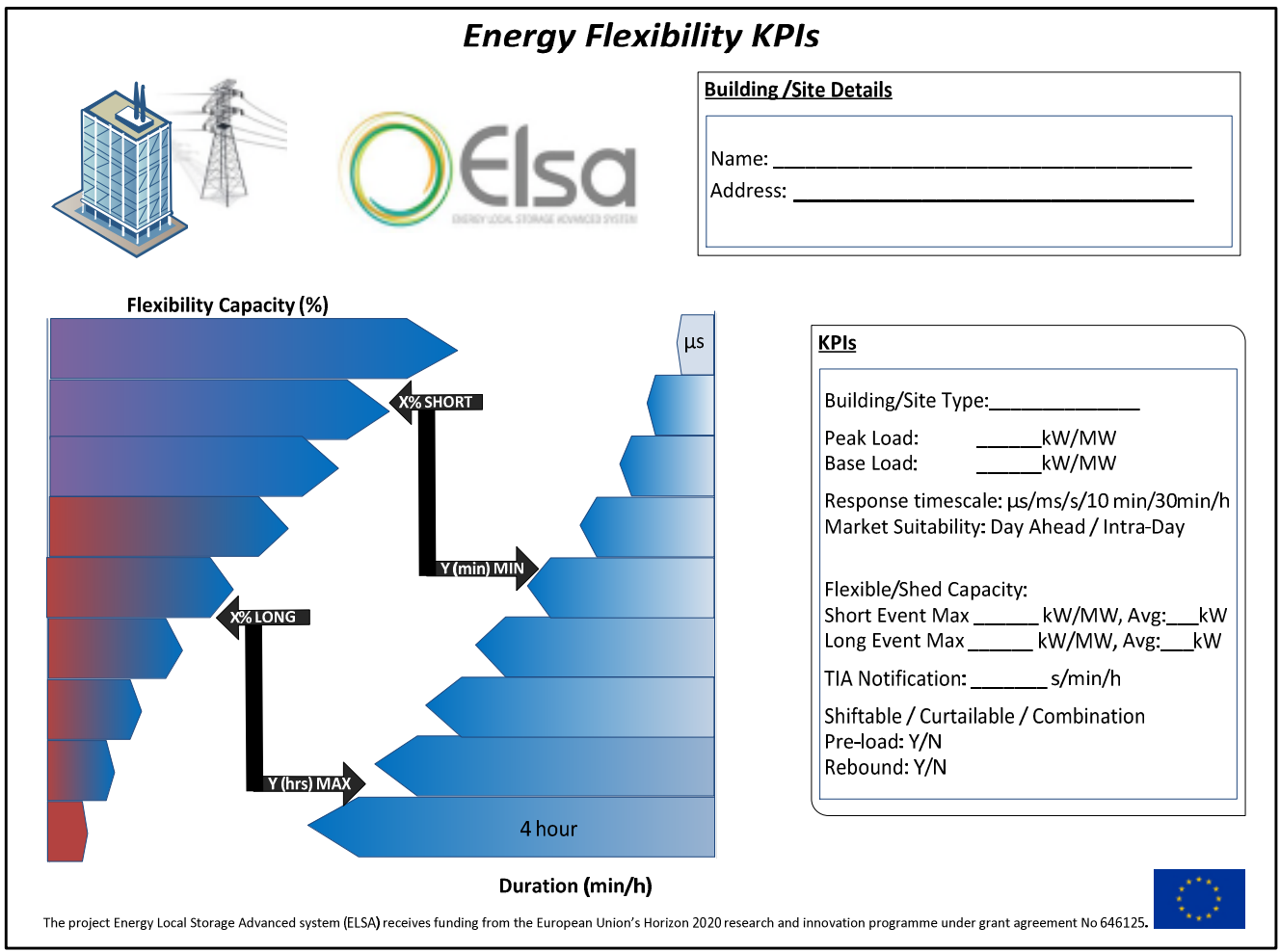

Fig. 2. KPI Label.

*C. Benchmarks*

To understand how much flexibility is typical, a number of other demonstration projects were reviewed. For comparison between projects, flexibility is denoted as a percentage of total site loads.

The Demand Response Research Centre (DRRC) at Lawrence Berkley National Laboratory (LBNL) in the US conducted a number of studies demonstrating flexibility in buildings [16] [17]. The studies included below were for Critical Peak Pricing (CPP) events. Using pre-cooling, a flexibility of between 10 - 25% of peak load over a three hour duration was demonstrated [17]. Average savings of between 7 - 9 % with peak savings of up to 56 % over shorter durations, was achieved when automated demand response (ADR) was implemented in 28 buildings [16].

A more recent European demonstration [11] achieved a 15% load reduction when maximizing bid power over a half hour period. The demonstration involved 8 pilot sites, battery storage and a PV array. The pilot sites each provided a single load, heating in buildings or pumps in industrial sites. The PV and storage elements were similar to the pilot site presented here but were managed by an aggregator, instead of by a local energy management system as is proposed for this project.

*D. KPI Label*

The Key Performance Indicators (KPI) label proposed in Fig 2 gives a useful indication of the range of flexibility that a site has the capability of providing based on an off-line assessment. It incorporates the key factors in a simple, easily understood format which can be interpreted at a glance.

## VI. PILOT SITE EXAMPLE

The Gateshead College Skills Academy for Sustainable Manufacturing and Innovation (SASMI) is a 5,700 m$^2$ building consisting of classrooms, offices and workshops. It is located adjacent to the Nissan manufacturing facility in Sunderland, UK. The peak power load in the building is approximately 140 kW and the base load is between 20 kW and 40 kW. To illustrate the methodology, a load profile from one sample day in winter is used. This is an important use case as DSBR operates during winter only at present.

Energy flexibility in the SASMI building will be provided by the Nissan Leaf second life battery system, a 40 kWp PV array and electrical loads in the building. The loads in the building which have the capability to provide flexibility include HVAC loads such as Air Handling Units (AHUs), and a Variable Refrigerant Flow (VRF) heat pump system.

An ICT platform incorporating the BEMS is in the process of being installed on site. Control and optimization algorithms will manage the loads and storage on site to provide flexibility services to an aggregator or the grid, in response to signals sent using the OpenADR protocol [18]. The battery management system, being developed Bouygues in cooperation with Nissan and Renault, will communicate with the BEMS via web services. The pilot site architecture is shown in Figure 3.

Two sample scenarios for flexibility, applying the methodology developed in this paper, are presented. The first is for a one hour flexibility event, the second is a four hour flexibility event.

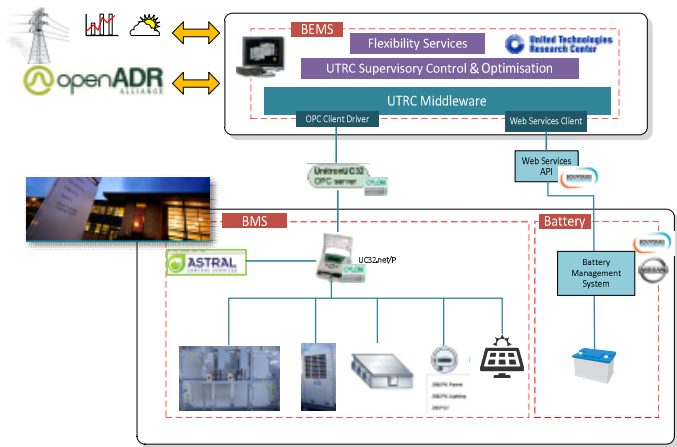

Fig. 3. Pilot Site System Architecture

*A. Scenario I: One Hour Flexibility Event*

The one hour flexibility event scenario, shown in Figure 4, illustrates the percentage of total load which the storage, on-site generation and loads have the capability to provide in response to a request such as demand response. If only the battery system was used, it could provide up to 26% flexibility. Adding PV output brings this up to 29% in winter and potentially 33% in summer. Estimated reductions of 10 % in HVAC load increases the flexibility to 32%.

It is worth noting that the heating system in the building is primarily gas fired, hence the low reduction in winter HVAC load.

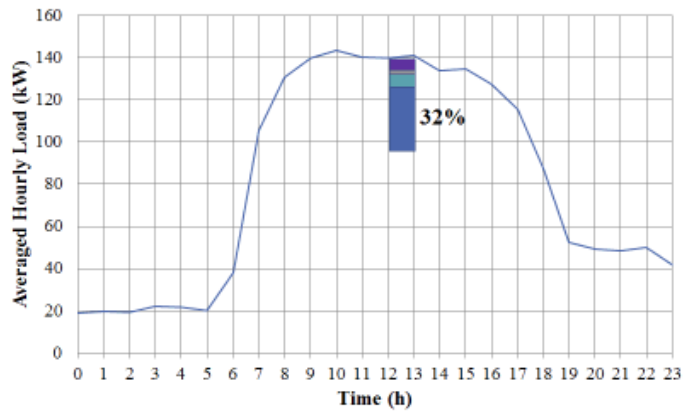

Fig. 4. One Hour Flexibility Event: PV, Battery & HVAC, Winter.

*B. Scenario II: Four Hour Flexibility Event*

The four hour flexibility event scenario, shown in Figure 5, illustrates the flexibility that may be achieved using the same sources over a longer time frame. The battery system can achieve 8% flexibility by itself. Adding the PV output, this increases to a peak of 11% in winter and potentially 18% in summer. With a 10% reduction in HVAC load, flexibility of up to 15% in winter may be achieved.

The contribution of HVAC to flexibility is much greater for the four hour event as the battery system capacity is spread over a longer timeframe, reducing its impact. Simulation or functional tests are required to determine a more precise figure for HVAC load reduction. The percentages included here are based on typical reductions in literature [16].

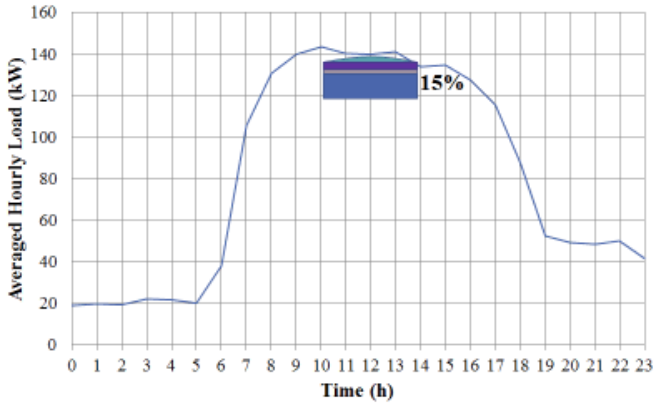

Fig. 5. Four Hour Flexibility Event: PV, Battery & HVAC, Winter.

Benchmark comparison against the data presented in Section V is favorable, as the site has greater than average flexibility and is within the largest maximum range.

TABLE I. BENCHMARK COMPARISON

| Benchmark 1 [17] | Benchmark 2 [11] | Site Flexibility (%) | Duration (h) |
|---|---|---|---|
| Avg 7 - 9% | Min ~ 7% | 8% - 15% | 4 h |
| Max 28 – 56% | Max ~18% | Max 32% | 1 h |

## VII. CONCLUSIONS

A methodology for assessing site flexibility for demand side services has been developed. The methodology is based on formulations for flexibility and defined in terms of optimisation input parameters. The categorisation approach applies these to create a standardised site assessment methodology, incorporating ISO 50002:2014.

The pilot site example demonstrated the application of the flexibility assessment and two scenarios were presented, one hour and four hour flexibility event. The flexibility capability of the site is between 8% and 32%, depending on the sources used and the duration of the event. These are within the range of the benchmark comparison.

Future work includes development of models for the system, completing the installation of the ICT system, development of algorithms to manage flexibility and validation at the pilot site.
ACKNOWLEDGMENT

This work was part funded by the European Commission under the Horizon 2020 program, ELSA project reference No. 646125. Sincere thanks to Zero Carbon Futures at Gateshead College for facilitating pilot site activities.